\documentclass[12pt,preprint]{aastex}
\newcommand{\beq}{\begin{equation}}
\newcommand{\eeq}{\end{equation}}
\newcommand{\beqn}{\begin{eqnarray}}
\newcommand{\eeqn}{\end{eqnarray}}

\newcommand{\sw}{{\rm sw}}

\begin{document}
\title{\bf{Clumps and Axisymmetric Features in Debris Discs} }
\author{Ing-Guey Jiang$^{1}$ and 
     Li-Chin Yeh$^{2}$}

\affil{
{$^{1}$ Department of Physics and Institute of Astronomy,}\\
{ National Tsing-Hua University, Hsin-Chu, Taiwan} \\
{$^{2}$ Department of Applied Mathematics,}\\
{ National Hsinchu University of Education, Hsin-Chu, Taiwan}\\     
}
\email{jiang@phys.nthu.edu.tw}

\begin{abstract}

This paper studied the structures of debris discs, focusing on the conditions that can form an axisymmetric-looking outer disc from systems with inner clumps. The main conclusion was that as long as the dominated dust grains are smaller than the blowout size, it is easy to form an axisymmetric-looking outer debris disc, which is part of a quasi-steady state of the whole system. This quasi-steady state is established through the balance between grain generations and a continuous out-going grain flow. Assuming there is an event that starts planetesimal collisions and the corresponding grain generations, this balance can be approached in a few thousand years. This result suggested that a quasi-steady-state picture could solve the possible mass budget problem of Vega's outer debris disc.

\end{abstract}


\newpage
\section{Introduction}

Through the detection of infrared excesses, circumstellar dust grains have been confirmed to exist around numerous main sequence stars. An interesting question is whether these dust grains are primordial or otherwise generated after the formation of star-disc systems. It could be that, for younger systems, a great fraction of dust grains are primordial. In fact, numerous surveys by the Spitzer Space Telescope have studied large numbers of samples of primordial discs, and have further classified them into {\it pre-transitional, transitional, evolved, and homologously depleted} discs (Carpenter et al. 2009). For older systems, the gas has already been depleted and the disc is mainly composed of dust grains and planetesimals. In such a gas-poor environment, the radiation pressure and collisional processes can remove dust grains on timescales much shorter than the stellar age. Thus, the dust must be replenished continuously through the collisions of planetesimals, and this is why the debris interpretation was proposed. 

Debris discs are gas-poor circumstellar discs with dust grains and planetesimals. Recent resolved images of debris discs have revealed complex structures with rings, warps and clumps (see the review in Wyatt 2008). It is unclear how these structures were formed. Theoretical works have shown that planetary companions might explain some of these structures. For example, Mouillet et al. (1997) tried to explain the warp in the beta Pictoris disc by a planet moving in an inclined orbit, and Wyatt (2003) used the resonant capture by a migrating planet to explain the clumps in the Vega system. 

Indeed, the discoveries of extra-solar planets (exoplanets) have made the above theories more attractive than other alternatives. In fact, the fast growing number of detected exoplanets has been making fundamental impacts on the related astronomical research. Since there are now more than 500 exoplanets, interesting statistics results have been produced by several groups (see Tabachnik \& Tremaine 2002, Zucker \& Mazeh 2002, Marchi 2007, Jiang et al. 2006, 2007, 2009, 2010). Moreover, dynamical studies on proto-planetary disc systems (see Takeuchi \& Artymowicz 2001, Takeuchi \& Lin 2002, Jiang et al. 2003, Zakamska \& Tremaine 2004, Jiang \& Yeh 2004a, 2004b, 2004c, 2007) could have good implications on the investigation of debris discs.

On the other hand, the formation and evolution of debris discs has become an important subject itself, because it plays a role in connecting the evolution of stars with planetary formation. It has thus attracted numerous investigations (see Thebault \& Augereau 2007, Grigorieva et al. 2007, Wyatt et al. 2007, Thebault \& Wu 2008, Stark \& Kuchner 2009). 

The general picture of a debris disc is as follows. There shall exist a region with many planetesimals colliding and continuously producing new dust grains. This is usually called the birth-ring region. Depending on the environment in the birth ring, certain amounts of grains with different size ranges can be produced. Naturally, the grains that are larger than the blowout size (defined as a grain size just on the boundary between being gravitationally bound and unbound) will stay around the system, and those smaller than the blowout size will be blown out. Therefore, whether a steady state can exist or not will depend on the size distribution of new dust grains, and also the grain production rate. If the planetesimal collision frequency and the grain production rate are large, and the new grains are dominated by large grains, the total mass in the debris disc will increase, until the collisional frequency goes down, or the produced grains become smaller. If the collision frequency is small, and the new grains are dominated by small grains, the total mass of the disc will decrease until a new balance is established, for example, by a stirring event that increases the collision frequency and produces more grains. 

A possible example of such a balance is the Vega system. Through the Spitzer data, Su et al. (2005) showed that the grains in the Vega system are dominated by those smaller than the blow-out size, and the disc follows a $1/R$ density profile, which implies a continuous out-going grain flow. However, the estimated grain loss rate is so high that it is unlikely to have been going on during the entire lifespan of Vega. Su et al. (2005) thus suggested that their observations have witnessed a large recent collisional event and its subsequent cascade. Another idea is to show that the Vega debris disc is actually dominated by larger long-lived grains (See Krivov et al. 2006, Muller et al. 2010), and that there is no mass budget problem. 

In order to further investigate this controversial and interesting problem, Jiang \& Yeh (2009) took the first step to study the existence of a self-consistent dynamical model of a debris disc with a continuous out-moving flow of grains. They found that a continuously replenishing model with new grains frequently generated from a ring region can exist. The realistic chemical compositions and grain size distributions are considered in the model. They also found that the necessary timescale to approach a steady state is only about 4000 years. 

However, the assumption by Jiang \& Yeh (2009) that all dust grains are generated homogeneously in an axisymmetric ring region needs to be relaxed, due to, as shown in many observations, the disc possibly being clumpy with complicated structures. For instance, Holland et al. (1998) and Wilner et al. (2002) showed that there are two clumps in Vega's inner disc. In contrast, Vega¡¦s outer disc is rather symmetric, as imaged by the Spitzer Space Telescope in Su et al. (2005). If collisional probabilities between planetesimals in clumpy regions are large, it is unclear how symmetric the outer disc could be, from a theoretical point of view. 

In order to study the conditions that new grains generated from the clumpy inner disc could form an axisymmetric outer disc, and also examine how long it would take to establish a symmetric quasi-steady-state outer disc, this study improved on old models by Jiang \& Yeh (2009) to construct a more realistic new model. The Vega system was used for modeling and simulation, as it has the necessary characteristics. The results are useful for understanding the evolution of debris discs in general. 

In this paper, two rotating clumpy regions were set up, and new grains were generated from there. The grain distribution and the disc's density profile were then determined. The new grains were assumed to have been created through the collisions between planetesimals, and thus, the amount of grains would depend on the collisional frequency. As the total mass of the planetesimals in clumpy regions determines the collisional frequency, different mass models were explored. The goal was to investigate the conditions in which the outer disc could look axisymmetric under a dynamical model where new dust grains are generated from two clumps with a continuously replenishing process. The details of the models are shown in \S 2, and the results are presented in \S 3. Finally, a conclusion is provided in \S 4.  

\section{The Model}

As mentioned in the previous section, Holland et al. (1998) and Wilner et al. (2002) detected two clumps with dust emission peaks in Vega's inner disc. Moreover, two clumpy regions associated with dust emission peaks were numerically modeled as dust grains captured into resonances in Fig. 2 of Wilner et al. (2002). Wyatt (2003) also used the scenario of resonant captures to produce two clumps with sizes of approximately 20 to 40 AU. Given the existence of two clumps with higher densities of dust grains, the goal in this study was to explain the axisymmetric outer Vega disc, as shown in Su et al. (2005).

The authors proposed that, in Vega's debris disc, a planet will capture numerous planetesimals in a 3:2 resonance. As presented in Fig. 8.4 of Murray \& Dermott (1999), two clumps are most likely due to 3:2 resonances for particles with eccentric orbits. These particles are the planetesimals, which spend more time around the regions of the two clumps, and so the number densities of planetesimals in the two clumps are higher. Collisions between planetesimals produce the dust grains in these two clumps as seen in observational results. Some of the newly generated dust grains might stay at the inner disc, while some smaller grains will be blown out by 
the central star's radiation pressure to form the outer disc.

To construct a simple model to investigate the dynamics of newly generated dust grains and the formation of outer debris discs, the planet, resonant planetesimals and resonant dust grains were not actually not included in our calculations. Rather, 200 non-resonant planetesimals moving with circular orbits around the inner disc were studied. When these planetesimals enter two clumpy regions, there are certain probabilities to produce collisional events. In the simulations, 200 dust grains would be added into the system whenever a collision occurred. The gravitational force from planetesimals on dust grains was not considered, as this study was only interested in those grains escaping the attraction of a particular planetesimal.

\subsection{The Units}

For the equations of motion, the unit of mass was $M_{\odot}$,  
the unit of length was AU, and the unit of time was in years. 
Thus, the gravitational constant was
$G=6.672\times 10^{-11} ({\rm m^3/kg\,sec^2})
=38.925({\rm (AU)^3/M_{\sun} year^2})$, and the light speed was 
$c=3\times 10^8{\rm  (m/sec)}=6.3\times 10^4 ({\rm AU/year})$.

\subsection{The Models of Planetesimals and Collisional Frequencies}
In the debris disc, there must be a number of planetesimals moving around the inner disc that occasionally collide to create new dust grains. This study assumed that due to resonant captures, there were two clumpy regions having higher densities of planetesimals. In reality, no matter if they are within the clumpy regions or not, any pair of planetesimals have a chance to collide, as long as they are in close proximity to each other. Due to the chance of collisions in the clumpy regions being much higher, this study ignored low-frequency collisions in non-clumpy regions. In the simulations, the 200 target planetesimals moved on circular orbits in the inner disc, and were uniformly distributed in a birth-ring region from 60 to 80 AU. At a given time interval, those located inside the clumpy regions would generate new dust grains through collisional events. For each collisional event, 200 representative dust grains were created and added into the system. The target planetesimals continued to move on their circular orbits.

An approximate estimation on the collisional probabilities is necessary to obtain the correct time interval. For a target planetesimal passing through a region of planetesimals characterized by a constant spatial density, the model proposed in Vedder (1996) provides an excellent way to obtain the collisional frequencies and time intervals. The model in Vedder (1996) was built up for the study on the interactions between asteroids, but it could also be used this study. It considers the spatial distribution of asteroids (or planetesimals) with respect to the target at a random instant of time, characterized by the Poisson distribution. The theory has been generalized to be in any dimension, but in the usual case of three spacial dimensions, the expected time between collisions, $T_c$, is given by
\beq
T_c = [\frac{3\pi b}{8v}] [\frac{b}{r_c}]^2, \label{eq:T_colli}
\eeq
where $v$ is the relative velocity of the collisional pair, 
$r_c$ is the characteristic 
collision distance, and $b$ is a scale parameter.
According to Vedder (1996), $b$ is related to the spatial number density of 
planetesimals $\delta$ as
\beq
\delta= \frac{3}{2} \pi^{-3/2} \Gamma(3/2) b^{-3}, \label{eq:delta_b}
\eeq
where $\Gamma( )$ is the usual gamma function.
In the simulated models,  
$r_c$ was set as the summation of the radii of two planetesimals.
 
An unseen planet was assumed to be moving on a circular orbit with
an orbital radius $R_p=54$ AU. Moreover, due to 3:2 resonant captures,  
there were two regions having higher densities of planetesimals.
These were called clumps, and were centered at (0,70) and (-70,0)
at the beginning of the simulation, respectively.
These two clumps were assumed to be circular with a radius of  10 AU.
Because both clumps were produced by the resonance with the planet, they
moved around the central star with the same angular speed as the planet, 
i.e. $\omega=v_c/R=\sqrt{\frac{Gm_0}{R_p}}/R_p$, where $R_p=54$. 
 
Assuming all planetesimals had the same radius of 5 km, the number of 
planetesimals was about $5.686 \times 10^{9}$  
for a standard model with 1 $M_{\oplus}$ of planetesimals.
If these planetesimals were located in a clump with a radius of 10 AU, 
thickness 1 AU, the spatial number density of planetesimals 
$\delta=1.81 \times 10^{7}/{ {\rm AU}^3}$. 

For a planetesimal moving on a circular orbit at $R=70$ AU, the Keplerian
velocity is $v_{\rm K}= \sqrt{{Gm_0}/{70}}$. This study used  $v=0.5 v_{\rm K}$
as the approximate estimation on the relative velocity
between two planetesimals in a collisional event.
Since the values of $v$, and  $\delta$ were known , and $r_c=10$ km, it was 
found that 
the expected time between collisions $T_c$ was about 
$6 \times 10^{6}$ years. Assuming
assume $0.2 M_{\oplus}$  planetesimals were located 
in the birth-ring region between 
$R=60$ and $R=80$AU of the inner disc, there would  be
about $1.2 \times 10^{9}$ planetesimals. Since this study only use 200 
particles 
to represent them, each particle represented $6 \times 10^{6}$ planetesimals.
Therefore, in the simulations, as long as a representative planetesimal had
entered one of the clumps, this planetesimal would have a collisional event
once a year for a model with $1 M_{\oplus}$ located in each clump 
and $0.2 M_{\oplus}$ in the birth-ring region between 60 and 80 AU.
This study used the variable $T_{sc}$ as the simulated $T_c$, and set
$T_{sc}=1$ year for the standard mass model.

From the above calculations, it is obvious that the total mass of planetesimals would directly influence the collisional frequencies and the evolution of disc structures. In order to investigate this effect, this paper presented four mass models. On the other hand, since different sizes of grains might be generated during the collisions, and the effect of grain sizes is also important, this study employed three ranges of grain sizes:
(A) small grains with radii from 
                 $a_{\rm min}$=1.02 to $a_{\rm max}$=2.36 $\mu$m, 
(B) middle-size grains with radii from 
                 $a_{\rm min}$=5.05 to $a_{\rm max}$=9.95 $\mu$m,
and (C)  larger grains with radii from 
              $a_{\rm min}$=12.02 to $a_{\rm max}$=20.96 $\mu$m.
The list of 12 models and the corresponding 
$T_{sc}$ in the simulations is shown in Table 1.\\
\\

\centerline{ {\bf Table 1} The Models} 
\begin{center}
\begin{tabular}{|c|c|c|c|c|c|}
\hline
Model&$a_{\rm min}$($\mu$m)&$a_{\rm max}$($\mu$m)
&$M_{\rm clump}$($M_{\oplus}$) 
&$M_{\rm ring}$($M_{\oplus}$) & $T_{sc}$(years) \\\hline
A1     & 1.02     & 2.36     & 1  &  1/5 & 1  \\ \hline
A2     & 1.02     & 2.36     & 1/2 &  1/10 & 4  \\ \hline
A3     & 1.02     & 2.36     & 1/4 &  1/20 & 16  \\ \hline
A4     & 1.02     & 2.36     & 1/8 &  1/40 & 64  \\ \hline
B1     & 5.05     & 9.95     & 1  &  1/5 & 1  \\ \hline
B2     & 5.05     & 9.95     & 1/2 &  1/10 & 4  \\ \hline
B3     & 5.05     & 9.95     & 1/4 &  1/20 & 16  \\ \hline
B4     & 5.05     & 9.95     & 1/8 &  1/40 & 64  \\ \hline
C1     & 12.02    & 20.96    & 1  &  1/5 & 1  \\ \hline
C2     & 12.02    & 20.96    & 1/2 &  1/10 & 4  \\ \hline
C3     & 12.02    & 20.96    & 1/4 &  1/20 & 16  \\ \hline
C4     & 12.02    & 20.96    & 1/8 &  1/40 & 64  \\ \hline
\end{tabular} 
\end{center}

\subsection{The Dust Grain Models} 

As we know from Moro-Martin et al.(2005),
a grain with a chemical composition of C400 and ${\rm MgFeSiO_4}$
could have the largest and smallest values of 
optical parameter $\beta$, respectively.
Thus, both C400 and ${\rm MgFeSiO_4}$ grains were considered, so that the 
effect of chemical compositions towards 
two possible extremes could be completely 
included in the simulations.

For each collision between planetesimals, 100 C400 grains and 
100 ${\rm MgFeSiO_4}$ grains will be produced. These two kinds of dust grains
follow the same power-law size distribution, i.e.
\beq
\frac{dN}{da} = C a^{-3.5},
\eeq
where $N$ is the grain number, $a$ is the grain radius and, $C$ is a constant.
Numerically, it can be written as
\beq
\triangle N = C a^{-3.5} \triangle a,\label{eq:gsize1}
\eeq
where $\triangle a$ is the bin size and $\triangle N$
is the expected grain number in the bin with a grain size around $a$.
As in Jiang \& Yeh (2009), the bin size was set to be 
uniform in a logarithmic space, with $h=0.038286$.
In order to investigate the effect of grain sizes on the disc structure, 
three ranges of grain sizes were employed.
This paper set $a_{A}=1\mu$m, $a_{B}=5\mu$m, and $a_{C}=12\mu$m.
For Model A1, A2, A3, A4, this paper set 
$a_i=a_{A}\exp\{h\times (i-1)\}$ for $i=1,2,\cdots, 24$ and defined 
$\bar{a_i}=(a_{i+1}+a_{i})/2$ for $i=1,2,\cdots, 23$ 
as the possible grain sizes.
Using   Eq.(4),
 $$\triangle N_i(\bar{a_i})=C\bar{a_i}^{-3.5}(a_{i+1}-a_{i}) 
\quad {\rm for}\  i= 1,2,\cdots,23.$$
Using the above, produces
\beq
N_{\rm tot}\equiv\sum_{i=1}^{i=23}\triangle N_i(\bar{a_i})=C \sum_{i=1}^{i=23}
\bar{a_i}^{-3.5}(a_{i+1}-a_{i}).\label{eq:c}
\eeq
The total particle number $N_{\rm tot}=100$, and parameter $C$
could  be determined from Eq.(5).
Thus,
the first grain size was $\bar{a_1}$, and 
the number of grains with this size was 
${\rm INT}[\triangle N_1(\bar{a_1})]+1
={\rm INT}[C\bar{a_1}^{-3.5}(a_2-a_1)]+1$,
where INT is an operator to take the integer part of a real number.
The rest of the process was the same as in Jiang \& Yeh (2009).
Moreover, for Model B1-B4,  only replace $a_{A}$ need to be replaced by $a_{B}$, 
and similarly, for Models C1-C4, $a_{C}$ was used instead of $a_{A}$.

A total of 200 dust grains were added into the system to simulate each collision event between planetesimals and the corresponding grain generation. Initially, these 200 dust grains were distributed from $R=0.1$ AU to 1 AU from the target planetesimal's center, and the surface number density followed an 1/R function. These dust grains' initial velocities were the same as 
the planetesimal's at the time of the collisional event. 
All dust grains were assumed to be in a two-dimensional plane, 
i.e. the orbital plane of the planetesimals, and were governed by the 
gravity and radiation pressure from the central star. For any given time, 
the central star was fixed at the origin. 
The dust grain's equations of motion were: 

\beq\left\{
\begin{array}{ll}
&\frac{d^2 x}{dt^2}=-\frac{Gm_0(1-\beta)}{R^3}x-\frac{\beta_\sw}{c}
\frac{Gm_0}{R^2}\left[\left(\frac{\dot{R}}{R}\right)x+\frac{dx}{dt}\right],\\
&\frac{d^2 y}{dt^2}=-\frac{Gm_0(1-\beta)}{R^3}y-\frac{\beta_\sw}{c}
\frac{Gm_0}{R^2}\left[\left(\frac{\dot{R}}{R}\right)y+\frac{dy}{dt}\right],\\
\end{array}\right.\label{eq:ini1}
\eeq
where,
\beq\left\{
\begin{array}{ll}
&R=\sqrt{x^2+y^2}, \\
&\dot{R}=\frac{x}{R}\frac{dx}{dt}+\frac{y}{R}\frac{dy}{dt},\\
&\beta_\sw=(1+\sw)\beta,
\end{array}\right.\label{eq:dot_r}
\eeq 
further, $(x,y)$ is the coordinate of a particular dust grain, 
$G$ is the gravitational constant, $m_0$ is the central star's mass
(2.5$M_{\sun}$), 
$c$ is the speed of light, $\beta$ is the ratio between the 
radiation pressure force and the gravitational force,
and ``sw'' is the ratio of the solar wind drag 
to the P-R drag. In this paper, ${\rm sw}$ was taken to be zero.
NAG Fortran Library 
Routine D02CJF was used  to numerically solve the 
above set of equations of motion.
The Mie scattering theory and Vega's spectrum (see Appendix A 
of Jiang \& Yeh 2009) were employed 
in the calculations of the optical parameter 
$\beta$ for dust grains.
The blowout sizes were the grain sizes with optical parameter 
$\beta=0.5$ .  For C400 grains, this was when grain radius 
$a = 12.02 \mu$m,
and for ${\rm MgFeSiO_4}$ grains, it was when $a= 8.44\mu$m.

On the other hand, as can be seen from the equations of motion, 
the planetesimals' and the planet's influences were not included, 
as this paper only focused on those dust grains that were escaping 
from the planetesimals. Moreover, as shown in Jiang \& Yeh (2009), the planet was not important for the outer debris disc, which is composed with smaller grains, and therefore, this paper mainly studied the structure of the outer disc. 

Fig. 1(a) shows the grain size distributions for all of the models. The histograms of the $\beta$  values for both the C400 and ${\rm MgFeSiO_4}$ grains of these models are shown in Fig. 1(b) to (d). From these $\beta$ histograms, the fractions of grains that were larger or smaller than the blow-out sizes could be easily estimated. For Models A1-A4 (Fig. 1(b)), all of the dust grains were smaller than the blowout size. For Models B1-B4 (Fig. 1(c)), all of the C400 grains were smaller than the blowout size, and a small fraction of the ${\rm MgFeSiO_4}$ grains were larger than the blowout size. For Models C1-C4 (Fig. 1(d)), all of the ${\rm MgFeSiO_4}$ grains were larger than the blowout size, but some C400 grains were near the blowout size.

\section{The Simulations}

In this section, the dust grains distributions and debris disc structures for all of the models are described and discussed. Although computing facilities used in this research were able to do simulations up to a longer time-scale, the disc profiles were already approaching steady states at $t=1000$ years, and thus, the simulations shown in this section started at time  $t=t_0\equiv 0$ and terminated at $t=t_{\rm end}\equiv 1200$ years. As discussed in Jiang \& Yeh (2009), for such a short time-scale, the collisions between dust grains could be ignored.

In order to distinguish the evolutionary processes 
of long-lived and short-lived grains, the 
dust grains were divided into two groups according to their $\beta$; i.e. 
the larger grains were those with $\beta < 0.5$ and  
the smaller grains were those with $\beta \ge 0.5$.
In Fig. 2, the disc's surface mass density $S$ as a function of radius $R$
at $t=t_{\rm end}\equiv 1200$ are shown for all 12 models.
The crosses represent larger grains (those with $\beta < 0.5$)
and the triangles represent smaller grains (those with $\beta \ge 0.5$).
The circles represent the disc's surface mass density with all grains,
and the solid curve represent the $1/R$ fitting function determined 
using the method described in Jiang \& Yeh (2009). As all of the grains in Models A1-A4 were smaller grains, the triangles overlapped the circles in the left panels. As all of the grains in Models C1-C4 were larger grains, the crosses overlapped the circles in the right panels. The results indicated that the debris disc¡¦s profiles in Models A1-A4 could be approximately well fitted by the $1/R$ 
functions, but others could not. This is due to, for Models A1-A4, all grains being small enough to get blown out continuously, forming a steady out-going flow; however, for other models, many larger grains did not get blown out continuously. On the other hand, the planetesimal collisional frequencies did not affect the results significantly, as can be seen from top to down in Fig. 2. The only difference is that, comparing the left-bottom panel (Model A4) with the left-top panel (Model A1), the deviation from the $1/R$ function was slightly larger. Nevertheless, the collisional frequency was still high enough to generate new grains and to maintain the steady-state profile.

In order to have a comparison between models through different views,  
two-dimensional contours of the surface mass densities 
at $t=t_{\rm end}\equiv 1200$ were presented, as shown in Fig. 3. 
It shows details of the face-on view of the disc structures in all models.
 Due to only small grains existing in Models A1-A4, grains flew away and the discs extended to a distance larger than $R=1000$ AU, as shown in the left panels. Because both large and small grains existed in Models B1-B4, the discs had radii around $R=1000$ AU, as shown in the middle panels. Finally, only large grains existed in Models C1-C4, so the discs were small and had radii about $R=300$ AU, as shown in the right panels. On the other hand, for Models A1-A4 (from top to down), due to the larger collisional frequencies between planetesimals in Model A1, the grain density was higher, and small clumps with even higher densities existed, as shown in the top panel. For Models B1-B4, the average densities were also higher in the top panel, but clumpy structures were more obvious in the bottom panel. This might be due to, for Model B4, the accumulation of large grains and the larger density-gradient near the central region, as there were few grains generated. For Models C1-C4, there were more fine structures, as shown in the bottom panel. 

In order to simulate what could be seen through an instrument, this study performed a convolution on Fig. 3 using a two-dimensional Gaussian function with  $\sigma=150$ AU to obtain Fig. 4. This figure was an analog of the 160$\mu$m map on Vega, as shown in Fig. 4 of Su et al. (2005). All panels shown in Fig. 4 looked completely axisymmetric.

To understand the detailed evolution of debris discs, this study presented the complete simulation of Model A1 in Figs. 5-6. Panel (i) is a snapshot at 
$t=100\times$i, where i=1, 2, 3,..., 12. As shown in Fig. 5, at $t=100$, some grains had already moved outward from their birthplace to between 200 and 300 AU. Grains continued to move out quickly, and reached $R=1000$ AU at $t=400$. The distribution of grains followed an $1/R$ function at $t=600$ and stayed at this steady state until the end of simulations at $t=1200$. This proved that the steady state could be approached and maintained in such a short time, and that it was enough to set  $t_{\rm end}\equiv 1200$. The generation of new grains and the out-going flow of grains were able to balance each other and form a continuous out-moving grain flow. Fig. 6 shows the two-dimensional contour map of the surface mass density of the grains in 12 snapshots of Model A1. The panels in Fig. 6 show the corresponding snapshots with the same relative positions in Fig. 5. The disc was small at $t=100$, but clumpy structures were already there. The peak of the clumpy structure rotated around and formed spiral arms. The disc became larger while grains continued to move out. The structure almost settled at $t=600$, although the whole pattern continued to rotate.

  The snapshots of Model A4 are presented in Figs. 7-8. As can be seen from Fig. 7, dust grains were moving outward continuously. Gradually, the grain distribution followed an $1/R$ function around the time from $t=500$ to $t=600$, and stayed at this steady state until the end of simulations at $t=1200$. However, the deviation from the $1/R$ functions was larger than those in Model A1. When looking at the debris disc from a face-on view as in Fig. 8, it was clear there was a lot of empty space without grains, due to the smaller collisional frequency. The debris discs resembled combinations of numerous straight arms with empty spaces in between. This explains why the deviation from the $1/R$ functions was larger.

Moreover, face-on snapshots of Model B1 are shown in Fig. 9 and the ones of Model B4 are shown in Fig. 10. In Model B1, the disc became larger gradually and a large number of spiral arms became clear in the end. In Model B4, the grain densities were lower, but numerous clumpy structures showed up in the central region. Furthermore, Model C1 is presented in Fig. 11 and Model C4 is shown in Fig. 12. The debris discs were small in these two models, but there were more fine structures in Model C4. The first few panels of Fig. 8-12 show the avalanches, as studied in Grigorieva et al. (2007). Due to the continuous generation of new dust grains through collisions in the two clumps, many more avalanches and spiral arms came out, which finally formed an overall axisymmetric structure. 

To summarize, the accumulated number of collisional events as a function of time are plotted in Fig. 13(a). There are four straight lines with different slopes, which correspond to four constant collisional frequencies. The total mass of dust grains within $R=1500$ AU of the disc as a function of time are shown in Figs. 13(b)-(d) for Models A1-A4, Models B1-B4 and Models C1-C4. The curves approached flat lines with zero slopes in Fig. 13(b) at a time when the number of newly generated grains balanced with the ones that were moving out of the disc (the outer boundary was defined as $R=1500$ AU) in Models A1-A4. This plot re-confirmed the establishment of steady grain flows in debris discs. However, the grains¡¦ total mass continued to constantly increase in Figs. 13(c)-(d) due to the accumulation of larger grains. The total mass went to a higher value in Fig. 13(d), as the grain sizes were larger in Models C1-C4. The flat line and the corresponding balance confirmed the establishment of a quasi-steady state. If these grains were assumed to represent the grains in Vega's debris disc,
their total mass would 
be scaled to about $3\times 10^{-3} M_{\oplus}$. 
As the total mass of the planetesimals in the models were about 0.1 to 1 Earth Mass, it would be possible to produce and maintain a 
$3\times 10^{-3} M_{\oplus}$ d debris disc for up to ten thousand years.

\section{Concluding Remarks and Implications}

Employing Vega as the example system, this paper studied the formation and evolution of the structures of outer debris discs through dynamical simulations. In a model with two clumps created by planetary resonant captures, the process of planetesimal collisions with dust grain generation was considered, and the 
grains' orbits and density distributions were determined. This paper focused on the effects of collisional frequencies and grain sizes on disc structures. Three grain size distributions and four planetesimal collisional frequencies are explored, leading to 12 models of simulations, as listed in Table 1. 

The simulations showed that planetesimals with a total mass of around one Earth Mass would have enough collisional frequency to produce a debris disc with a mass of about 0.001 Earth Mass, in about 1000 years. The collisional frequency was also high enough to erase the spiral arms and form an axisymmetric dust disc. This result implied that there was no hindrance to forming a continuous out-moving grain outflow, as the necessary total mass of the planetesimals and the timescale were both small. 

The main conclusions derived from the above models were that as long as the grains generated from the clumpy regions were small enough, it would be easy to form an axisymmetric-looking outer debris disc, and this quasi-steady state could be approached in about thousand years. In this paper, newly generated dust grains from clumpy regions created by planetary resonant captures successfully connected the observed clumpy inner disc with the smooth featureless outer debris disc, and the grain size was consistent with those well fitted by the Spitzer data as studied in Su et al. (2005) 

Su et al. (2005) estimated the total mass of Vega debris disc to be about  $3\times 10^{-3} M_{\oplus}$,  and as their data favored an interpretation of small grains, the lifetime was only about 1000 years. If this were a steady state established at the beginning of Vega's life, it would need a massive asteroidal reservoir. Therefore, Su et al. (2005) went on to suggest that it is due to a recent large collisional event (however, see Krivov et al. 2006 
and Muller et al. 2010 for another suggestion).

This paper proposed a quasi-steady-state picture instead of a sudden collision, and the results showed that it would only take about a thousand years to approach a quasi-steady-state with a continuous out-going flow. Moreover, as the dominated grain sizes depend on the details of the planetesimal collisions, three ranges of grain sizes were used in this paper. Due to gentle collisions having higher frequencies, the dominated grain size could be smaller ones, as those used in Models A1-A4. For these kinds of grains, it would be easy to form an axisymmetric-looking outer debris disc, as observed. A single big collision might not produce an axisymmetric outer disc. 

This quasi-steady-state could have started after some kind of stirring up to ten thousand years ago, and not at the beginning of Vega's life; therefore, there was no mass budget problem. The occasional stirring at various stages, due to large planetesimals or planets, is highly possible. With such a conclusion that a quasi-steady state could be established in a short time, the current observed Vega outer debris disc could be regarded as a transient phenomenon when compared with the star's lifetime.

 
\section*{Acknowledgment}
The authors thank the referee, Francesco Marzari, for his useful remarks and suggestions that improved the presentation of this paper enormously. We are grateful to the National Center for High-performance Computing for its computer time and facilities. This work is supported in part by the National Science Council, Taiwan, under NSC 98-2112-M-007-006-MY2.

\clearpage
\begin{figure}[htbp]
\includegraphics[width=15.cm]{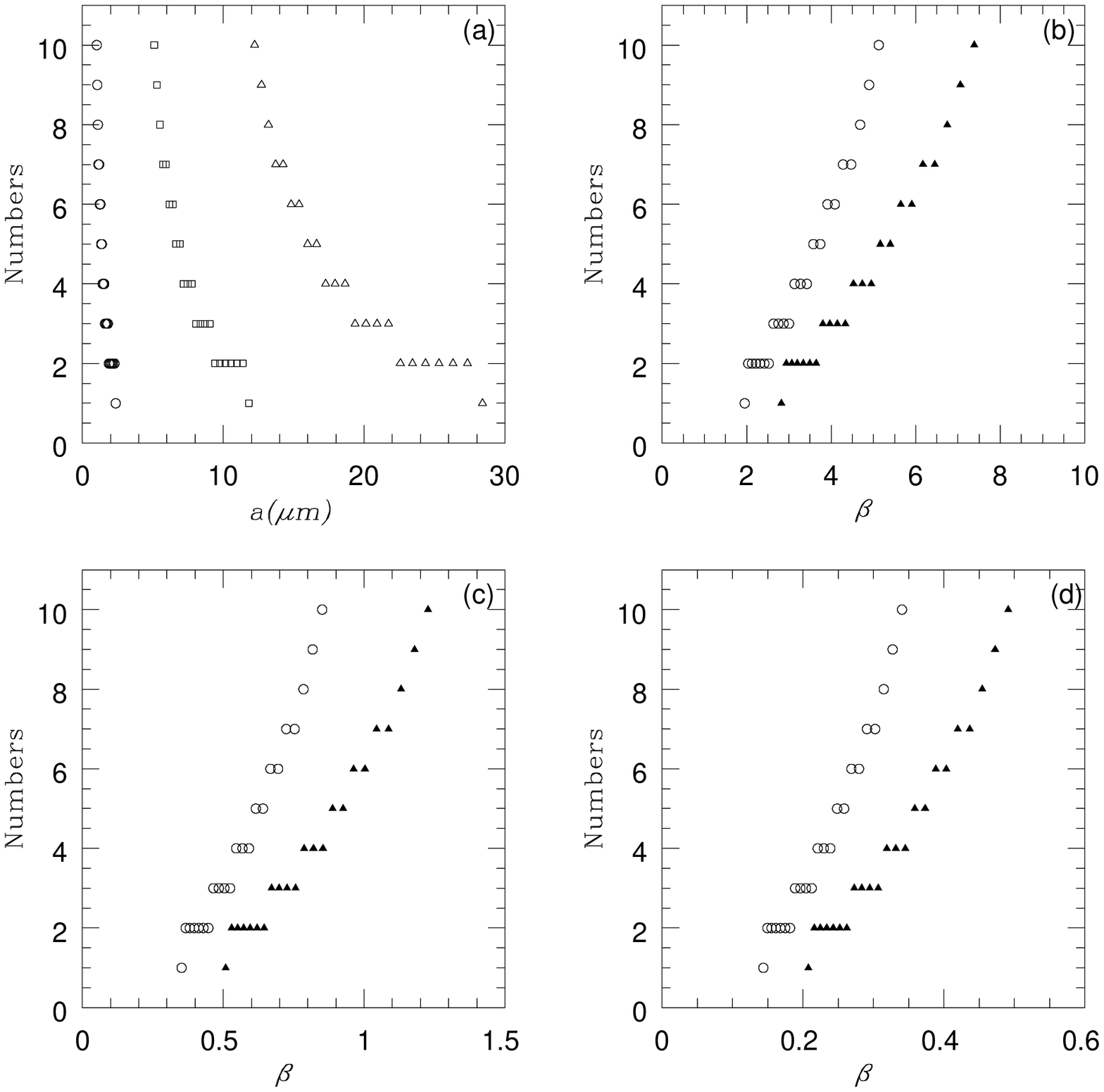}
\caption[]{(a) The grain size distributions, where the 
circles are for Model A1-A4, squares are for Model B1-B4,
triangles are for Model C1-C4.
(b) The histograms of the values $\beta$ of 
dust grains in Model A1-A4,
where full triangles are for C400 grains, 
circles are for ${\rm MgFeSiO_4}$ grains.
(c) The histograms of the values $\beta$ of 
dust grains in Model B1-B4,
where full triangles are for C400 grains, 
circles are for ${\rm MgFeSiO_4}$ grains.
(d) The histograms of the values $\beta$ of 
dust grains in Model C1-C4,
where full triangles are for C400 grains, 
circles are for ${\rm MgFeSiO_4}$ grains.
}
\end{figure}

\clearpage
\begin{figure}[htbp]
\includegraphics[width=15.cm]{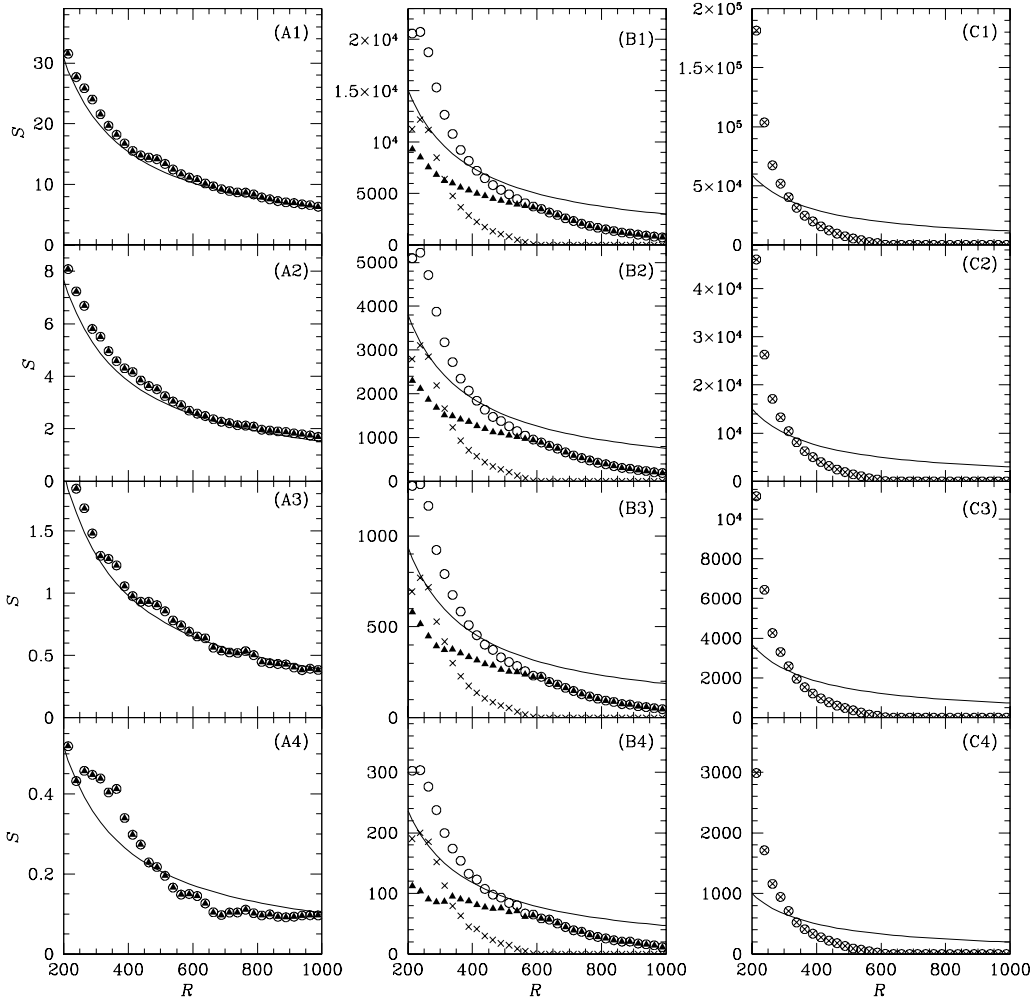}
\caption[]{The surface mass densities of grains 
as functions of radii at $t=t_{\rm end}\equiv 1200$ 
for all models.
The left panels are for Model A1-A4, middle panels
are for Model B1-B4, right panels are for Model C1-C4,
as indicated on the right-top corners of these panels.
In all panels, the triangles are for grains with
$\beta \geq 0.5$, the crosses are for grains with
$\beta < 0.5$, and the circles are the total. 
The solid curve is the best $1/R$ fitting function.
The unit of $R$ is AU and
the unit of $S$ is $10^{-12}{\rm g/AU^2}$. 
}
\end{figure}

\clearpage
\begin{figure}[htbp]
\includegraphics[width=15.cm]{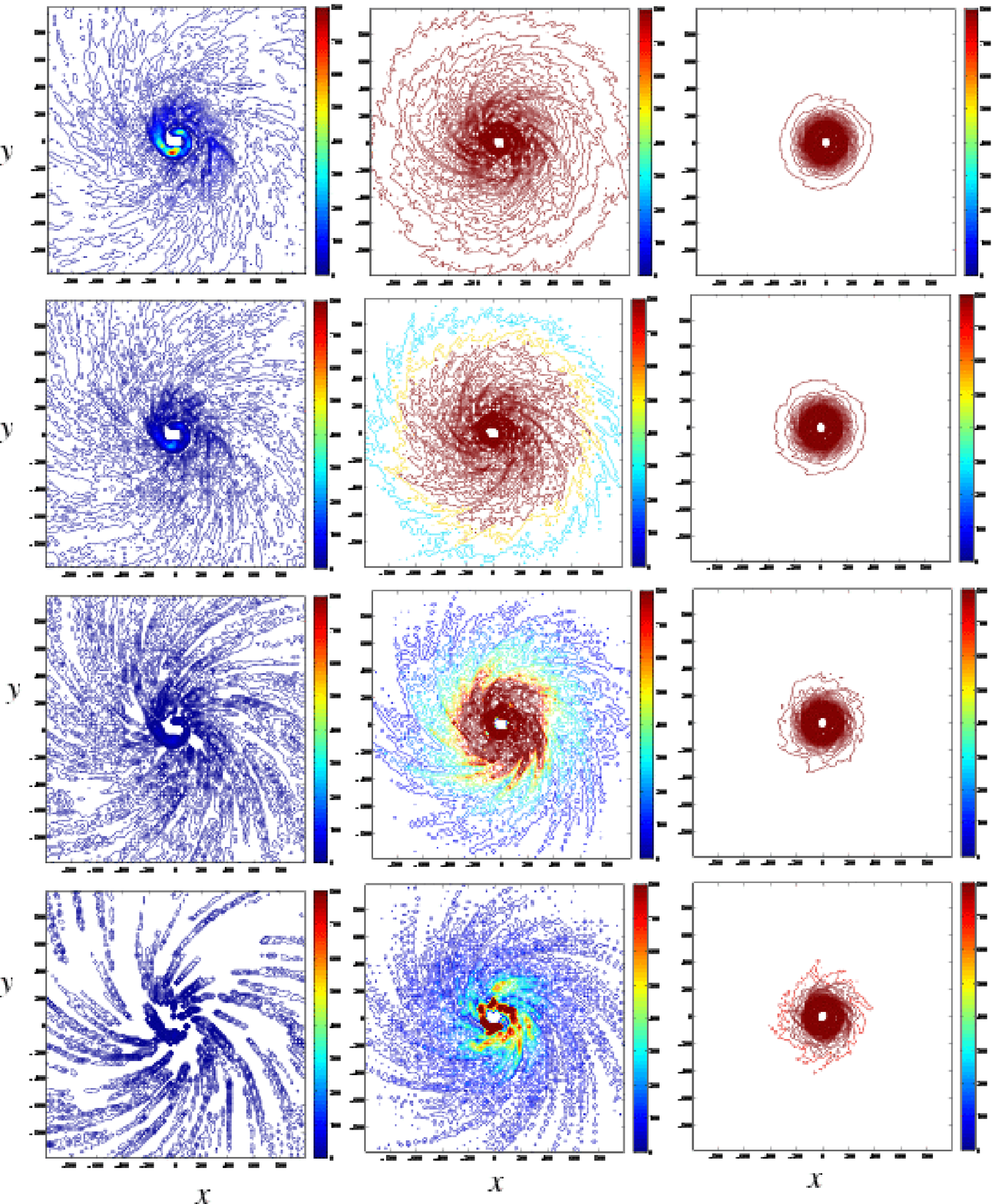}
\caption[]{The contours of surface mass densities of grains
at $t=t_{\rm end}\equiv 1200$ for all models.
The left panels are for Model A1-A4, middle panels
are for Model B1-B4, right panels are for Model C1-C4 
(from top to bottom). The unit of $x$ and $y$ is AU.
}
\end{figure}

\clearpage
\begin{figure}[htbp]
\includegraphics[width=15.cm]{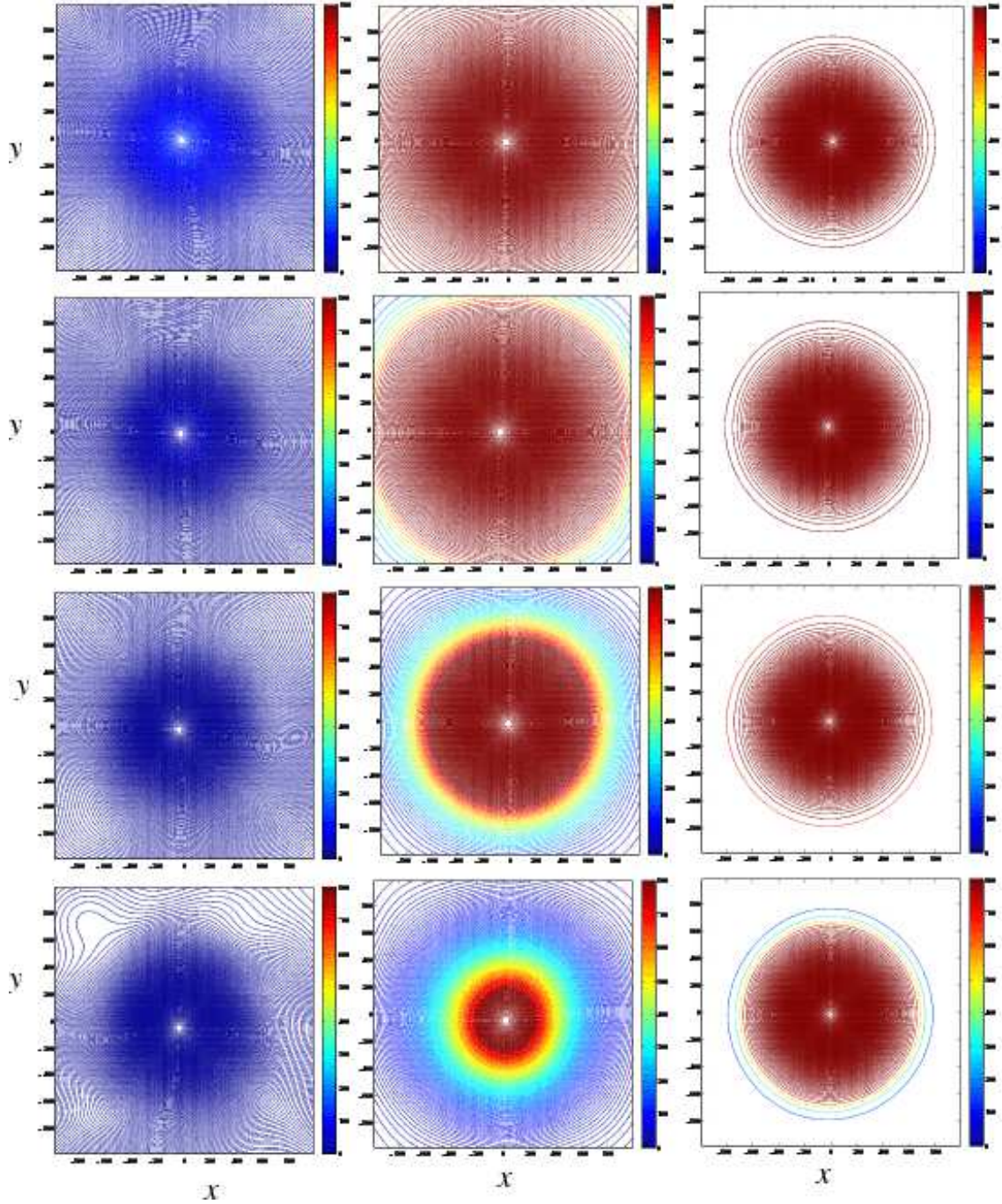}
\caption[]{The contours of surface mass densities of grains
at $t=t_{\rm end}\equiv 1200$ for all models, after the 
convolution by a Gaussian function (with $\sigma=150$ AU) is done.
The left panels are for Model A1-A4, middle panels
are for Model B1-B4, right panels are for Model C1-C4 
(from top to bottom). The unit of $x$ and $y$ is AU.
}
\end{figure}

\clearpage
\begin{figure}[htbp]
\includegraphics[width=15.cm]{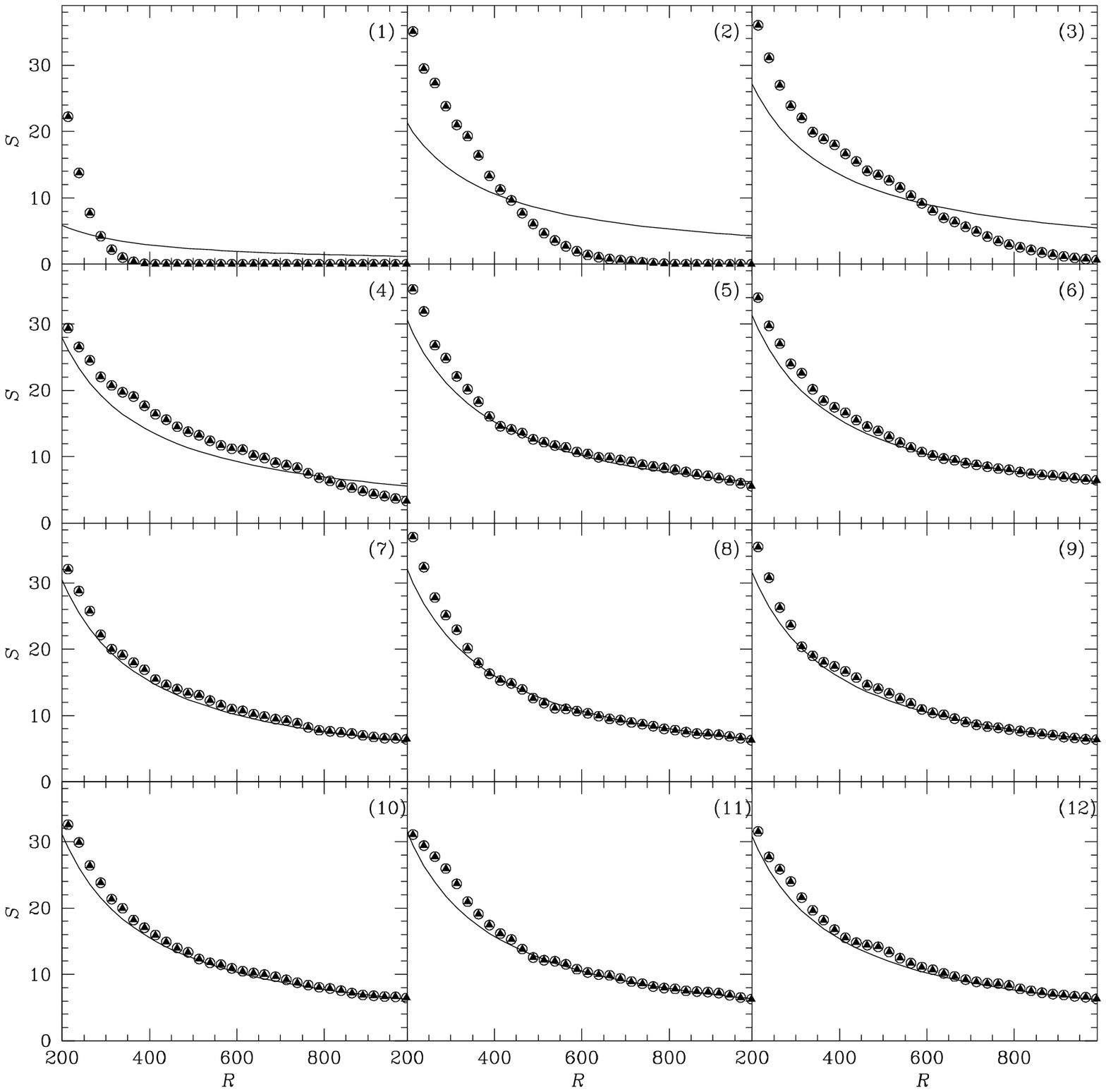}
\caption[]{The surface mass densities of grains 
as functions of radii for Model A1.
The ith panel ($i=1,2,..., 12$) is at
the time $t=100\times i$, where
the solid curve is the best $1/R$ fitting function.
In all panels, the triangles are for grains with
$\beta \geq 0.5$, the crosses are for grains with
$\beta < 0.5$, and the circles are the total. 
The unit of $R$ is AU and
the unit of $S$ is $10^{-12}{\rm g/AU^2}$. 
}
\end{figure}

\clearpage
\begin{figure}[htbp]
\includegraphics[width=15.cm]{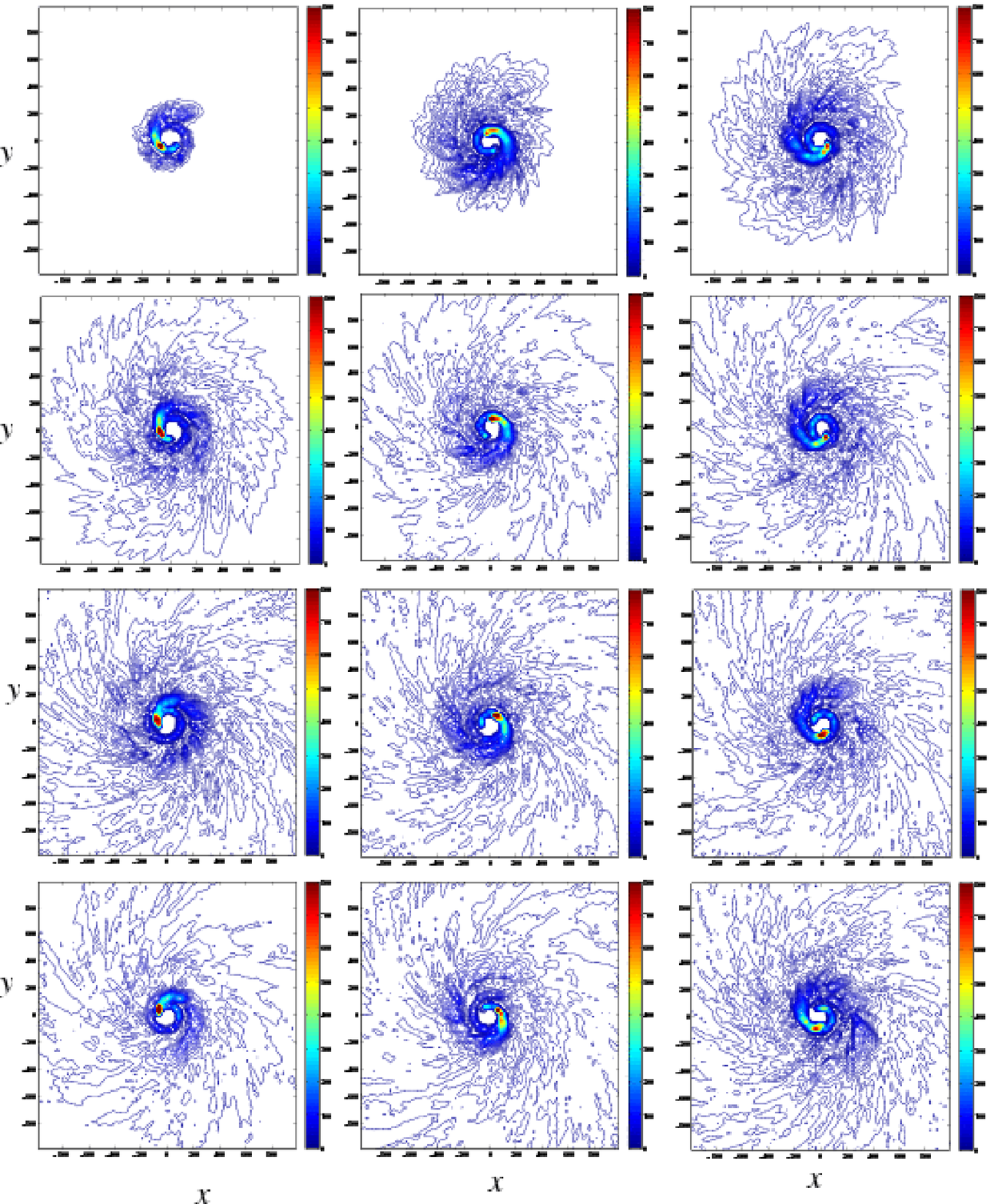}
\caption[]{The contours of surface mass densities of grains
on the $x-y$ plane for Model A1. The time for each panel is as 
for the corresponding panel with the same relative position in Fig. 6.
So, the ith panel ($i=1,2,..., 12$) is at
the time $t=100\times i$.
The unit of $x$ and $y$ is AU.
}
\end{figure}

\clearpage
\begin{figure}[htbp]
\includegraphics[width=15.cm]{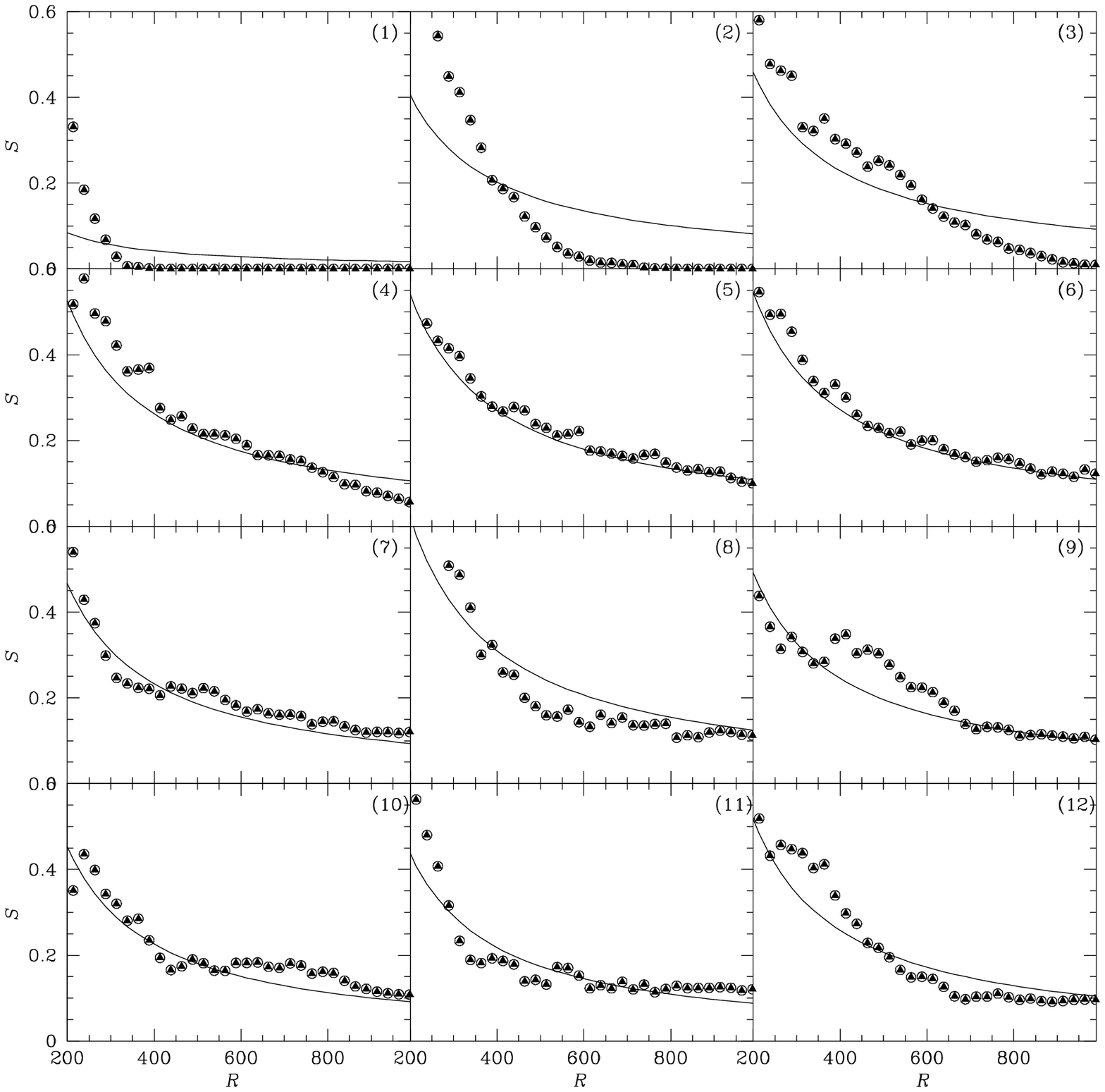}
\caption[]{The surface mass densities of grains 
as functions of radii for Model A4.
The ith panel ($i=1,2,..., 12$) is at
the time $t=100\times i$, where
the solid curve is the best $1/R$ fitting function.
In all panels, the triangles are for grains with
$\beta \geq 0.5$, the crosses are for grains with
$\beta < 0.5$, and the circles are the total. 
The unit of $R$ is AU and
the unit of $S$ is $10^{-12}{\rm g/AU^2}$. 
}
\end{figure}

\clearpage
\begin{figure}[htbp]
\includegraphics[width=15.cm]{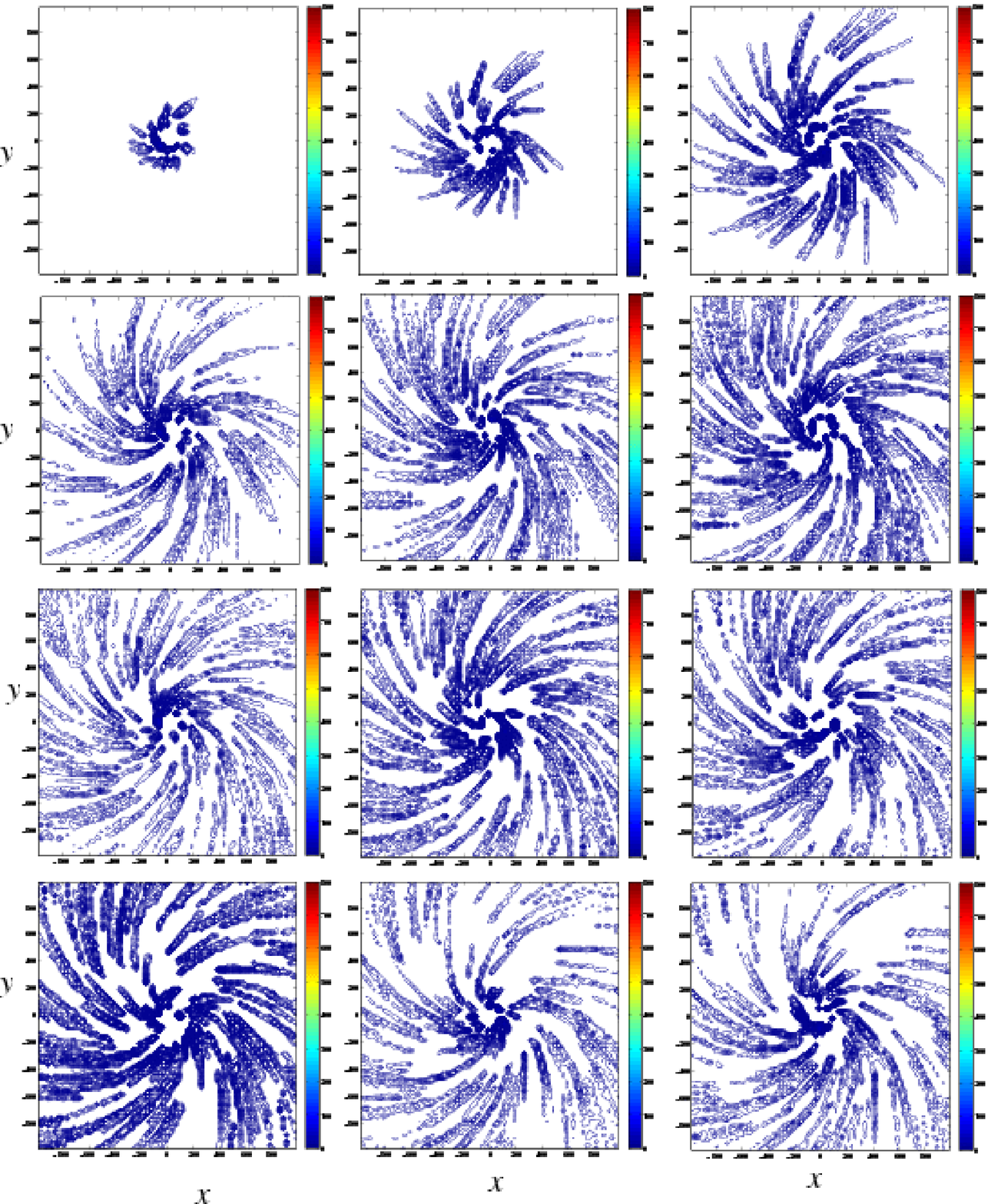}
\caption[]{The contours of surface mass densities of grains
on the $x-y$ plane for Model A4. The time for each panel is as 
for the corresponding panel with the same relative position in Fig. 8.
So, the ith panel ($i=1,2,..., 12$) is at
the time $t=100\times i$.
The unit of $x$ and $y$ is AU.
}
\end{figure}

\clearpage
\begin{figure}[htbp]
\includegraphics[width=15.cm]{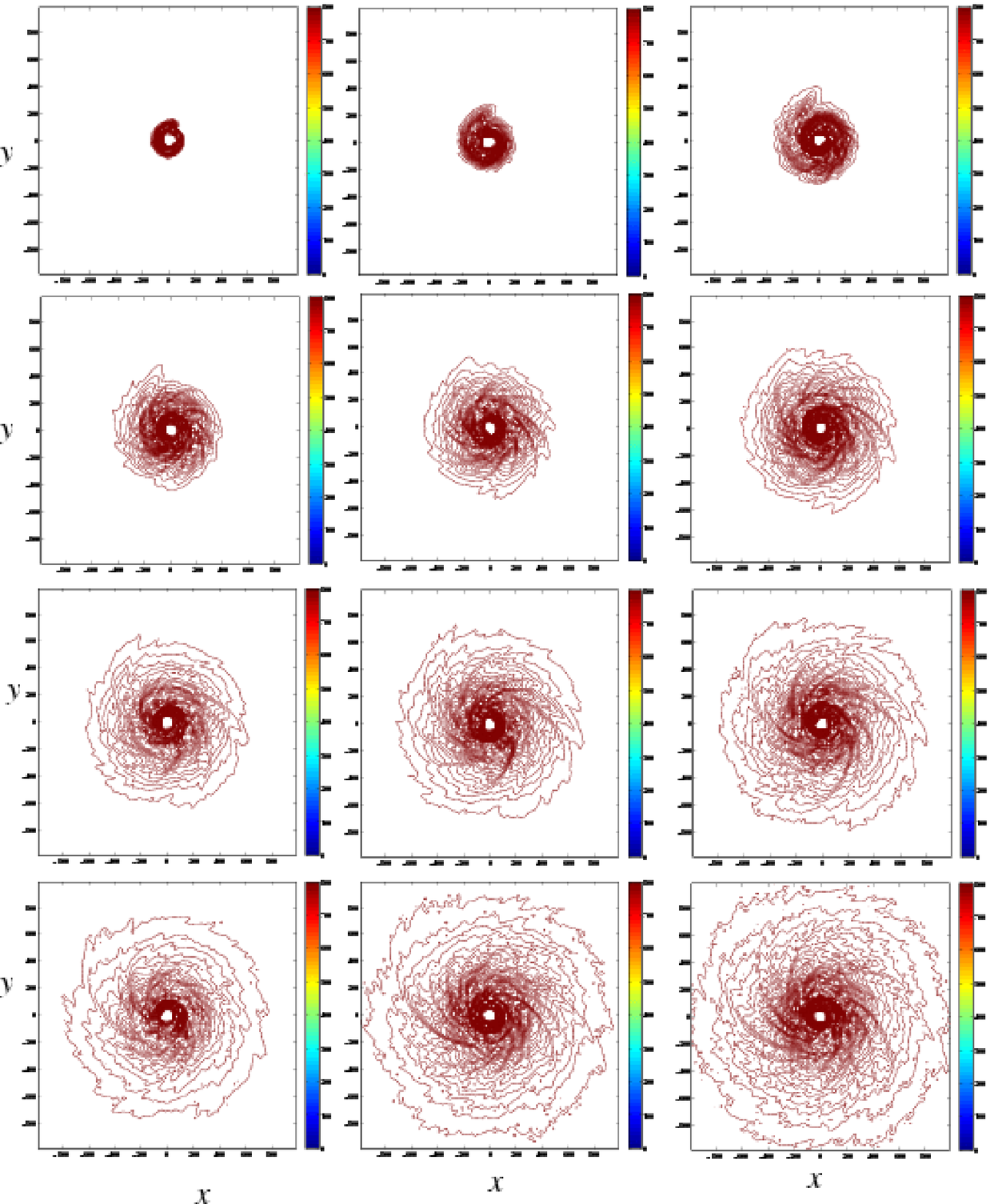}
\caption[]{The contours of surface mass densities of grains
on the $x-y$ plane for Model B1. The time for each panel is as 
for the corresponding panel with the same relative position in Fig. 8.
So, the ith panel ($i=1,2,..., 12$) is at
the time $t=100\times i$.
The unit of $x$ and $y$ is AU.
}
\end{figure}

\clearpage
\begin{figure}[htbp]
\includegraphics[width=15.cm]{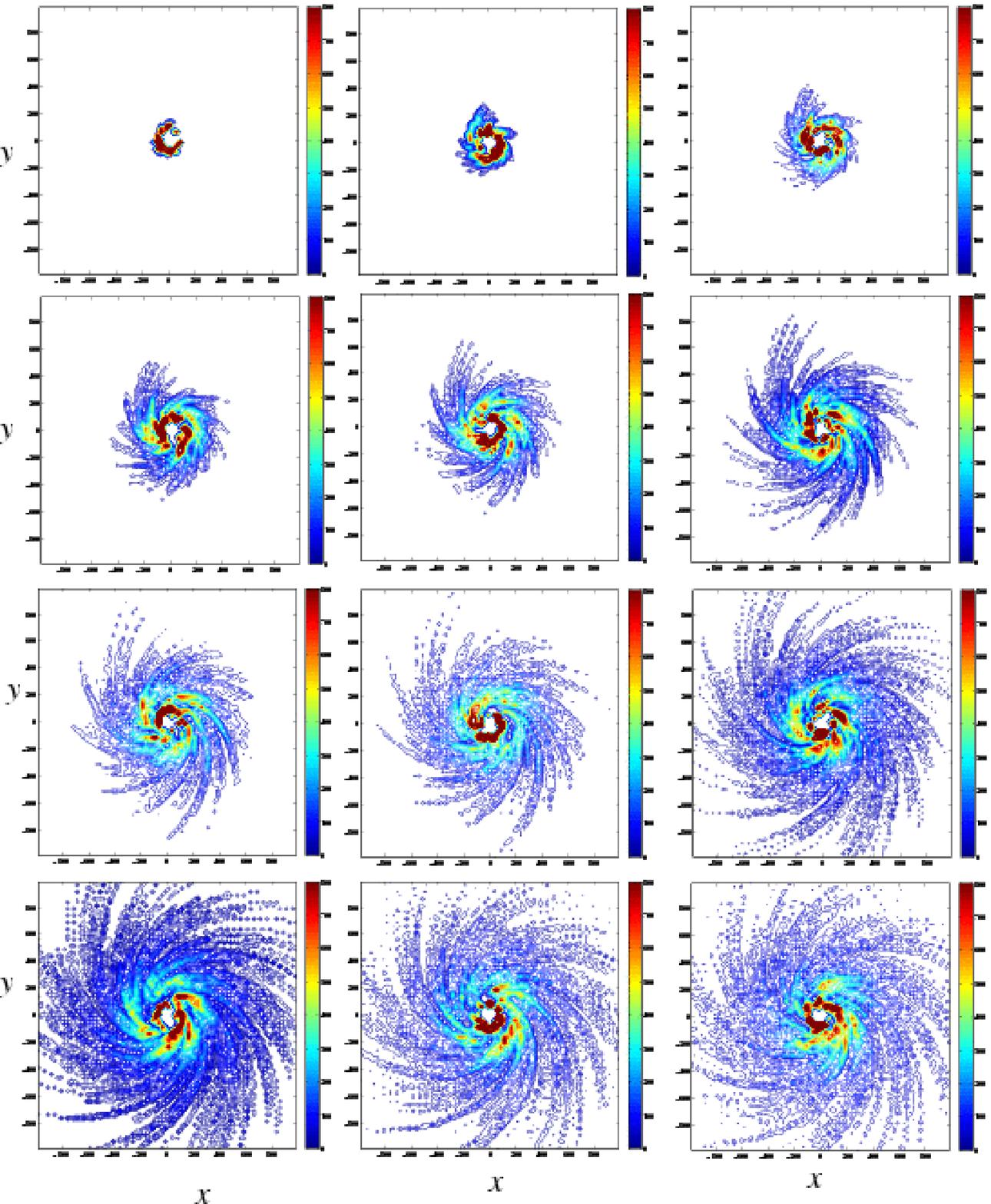}
\caption[]{The contours of surface mass densities of grains
on the $x-y$ plane for Model B4. The time for each panel is as 
for the corresponding panel with the same relative position in Fig. 8.
So, the ith panel ($i=1,2,..., 12$) is at
the time $t=100\times i$.
The unit of $x$ and $y$ is AU.
}
\end{figure}

\clearpage
\begin{figure}[htbp]
\includegraphics[width=15.cm]{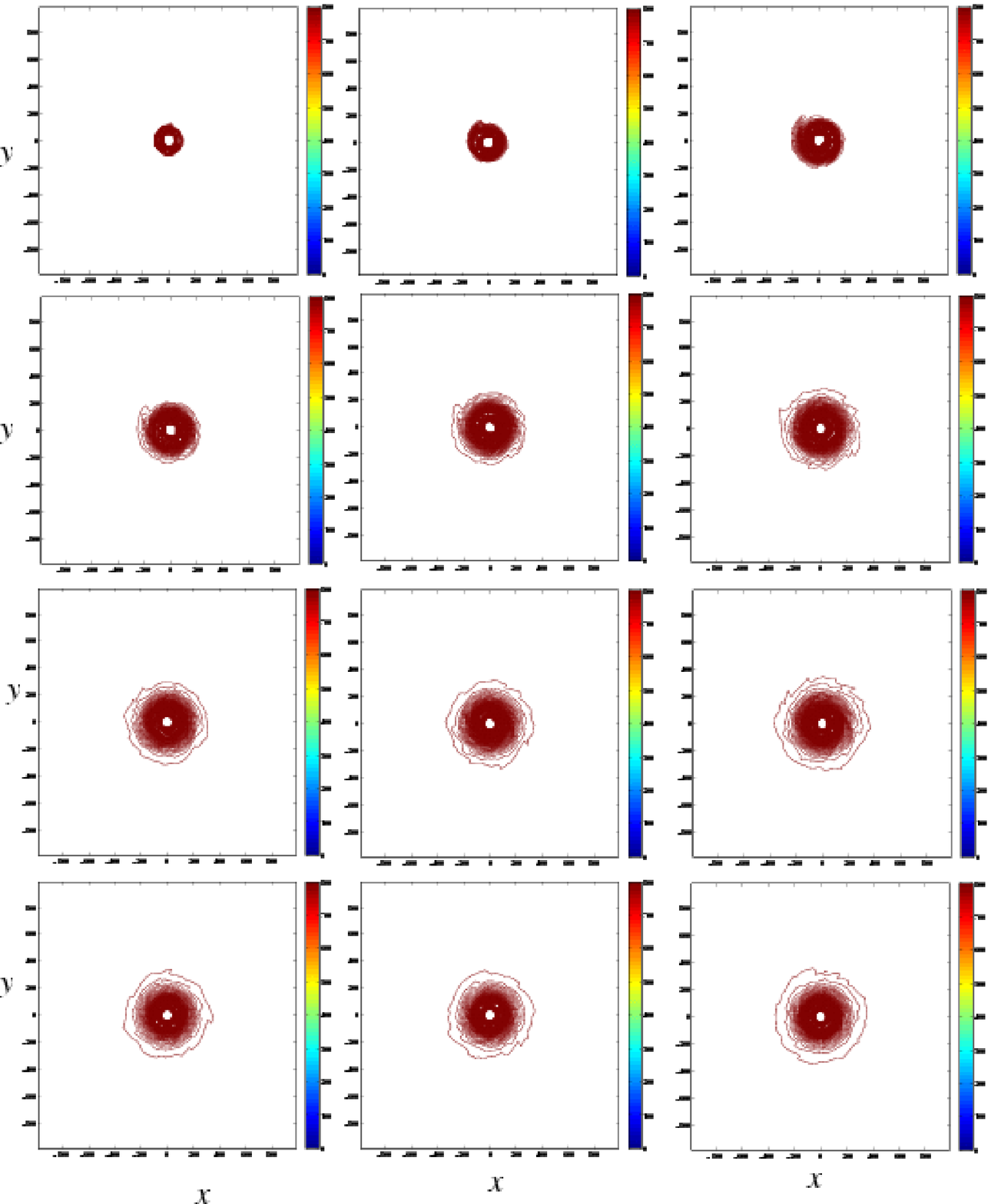}
\caption[]{The contours of surface mass densities of grains
on the $x-y$ plane for Model C1. The time for each panel is as 
for the corresponding panel with the same relative position in Fig. 8.
So, the ith panel ($i=1,2,..., 12$) is at
the time $t=100\times i$.
The unit of $x$ and $y$ is AU.
}
\end{figure}

\clearpage
\begin{figure}[htbp]
\includegraphics[width=15.cm]{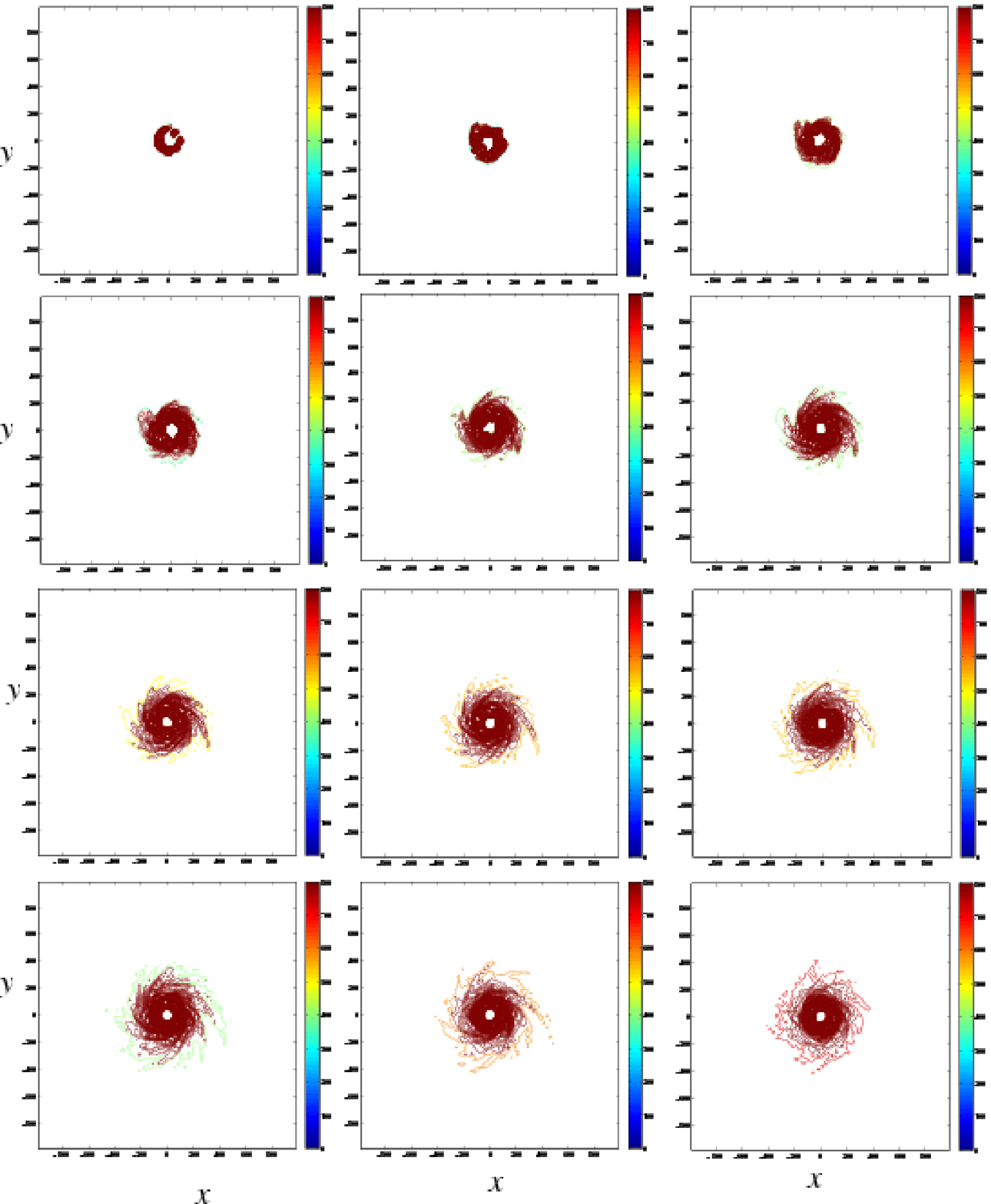}
\caption[]{The contours of surface mass densities of grains
on the $x-y$ plane for Model C4. The time for each panel is as 
for the corresponding panel with the same relative position in Fig. 8.
So, the ith panel ($i=1,2,..., 12$) is at
the time $t=100\times i$.
The unit of $x$ and $y$ is AU.
}
\end{figure}

\clearpage
\begin{figure}[htbp]
\includegraphics[width=15.cm]{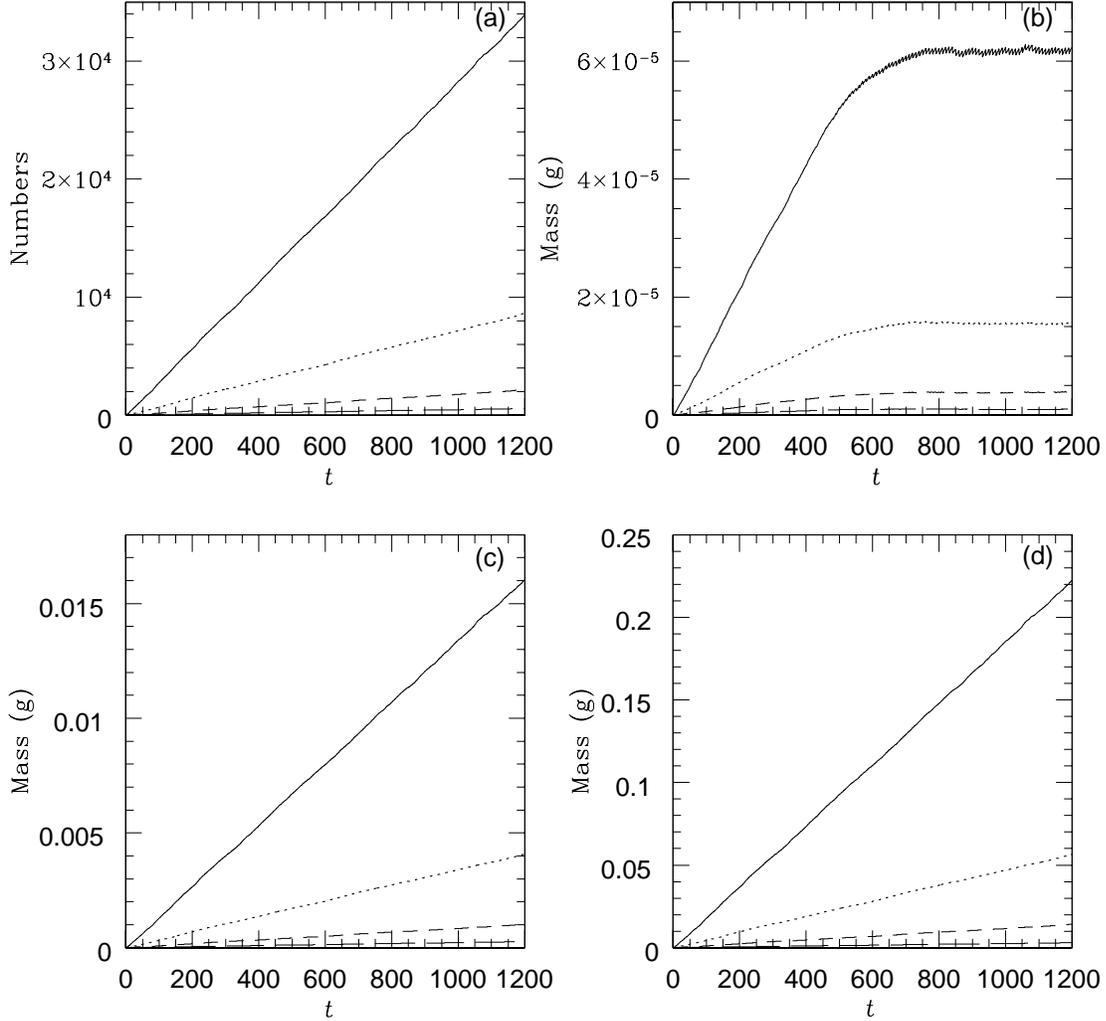}
\caption[]{
(a) The accumulated number of 
collisional events as a function of time. 
(b) The total mass of dust grains within $R=1500$ AU of the disc
as a function of time for Model A1-A4.
(c) The total mass of dust grains within $R=1500$ AU of the disc
as a function of time for Model B1-B4.
(d) The total mass of dust grains within $R=1500$ AU of the disc
as a function of time for Model C1-C4.
In all panels, the unit of time is year, and 
the solid curve is for $T_{sc}$= 1 year,
the dotted curve is for $T_{sc}$= 4 years,
the short  dashed curve is for  $T_{sc}$= 16 years,
the long dashed curve is for $T_{sc}$= 64 years.
}
\end{figure}

\end{document}